# Applications of Secure Multi-Party Computation in Financial Services

Brahim Khalil Sedraoui[1], Abdelmadjid Benmachiche[1], Amina Makhlouf[1], and Chaouki Chemam[1]

[1] University of Chadli Bendjedid, Faculty of Sciences & Technology, El Tarf, Algeria

**Abstract**

The concept of Secure Multi-Party Computation (SMPC) is a cryptographic service that allows generating analysis of sensitive data related to finance under the collaboration of all stakeholders without violating the privacy of the research participants. This article shows the increasing significance of privacy protection in the contemporary financial services, where various stakeholders should comply with stringent security and regulatory standards. It discusses the main issues of scalability, computational efficiency, and working with very large datasets, and it identifies the directions of future research to make SMPC protocols more practical and efficient. The results highlight the possibility of SMPC to facilitate safe, transparent, and trustful financial transactions in an ecosystem that is becoming more digital.

**Keywords**

Secure Multi-Party Computation (SMPC), data security, Privacy-preserving solutions.

## 1. Introduction to Secure Multi-Party Computation (SMPC)

Secure Multi-party computations (SMPC) allow multiple parties to compute a function over their private inputs without disclosing them to the other parties [1]. This aspect allows collaborating with untrusted parties and ensures the protection of confidentiality which makes SMPC especially relevant to such spheres as finance, healthcare, marketing, and e-learning platforms like Massive Open Online Courses (MOOCs) [2], [3] and Open Classrooms [6], [7], where confidential student or institutional information can be a mandatory condition. Since the introduction of SMPC in the late 1980s, two main directions of research have existed, one of which is the creation of effective and practically implementable solution, and the other is the development of protocols that are easy to use by non-experts [8]. Suggestions range to carefully optimized protocols that are task specific, to full-scale frameworks and libraries that can be used to support the transparent implementation of SMPC on generic applications.

In the simplest version of its formulation, SMPC allows a group of n participants to estimate functions based on their confidential inputs and reveal no information other than the final result. The existence of such feasibility assumes an honest majority, that is, that majority of those involved are loyal to the given protocol. The highest confidentiality is the case of semi-honest adversary model where partners do not gain any information about a counterpart but only that that is implied by the calculated result.

## 1.1. Definition and Key Concepts

SMPC protocols provide a strictly defined framework that has the capability to support collaborative computation on sensitive data [9]. Participants provide their own information and the protocol provides an output that is sent to everyone. A protocol is considered to be statistically secure when no data about the input of honest participants can be deduced, and even in the situation when corrupt parties participate.

SMPC families of protocols have been designed. The garbled circuit protocol designed by Yao allows parties to make or reveal encrypted values, ciphertexts, in the course of computation [10]. The protocols can also be categorized based on their communication complexity; e.g. bounded- round- based protocols place a limited number of messages exchanged to a bound of a given number. In the Common Reference String (CRS) model, a random string that is pre-defined and is available only to the participants is included in the computation [11].

SMPC protocols are resistant to passive adversaries, the players that follow the protocol but are interested in additional information, as well as malicious adversaries, who can randomly deviate; these attacks are normally countered by using zero-knowledge proofs. Recent advances have provided practical protocols which can withstand as many as t out of n actively corrupt parties, with computationally efficiency on the assumption that such properties as homomorphic encryption and secure point functions are available.

For example, in e-learning platforms including MOOCs [4], [5] and Open Classrooms [12], [13], the student information, including assessments, engagement indicators, and personal details, can require processing to be utilized in analytics or teamwork. SMPC helps to perform these calculations and protect the personal information allowing the servers to perform exactly those functions which are defined by the protocol [8]. While functions such as polynomial evaluation, inner products, maximum, matching, or sorting can be securely computed in semi-honest models, richer function classes may not allow complete privacy. he findings of Yao prove that it is not possible to hide a certain set of input in an arbitrary set of functions that can be computed in a non-real-time when one of the parties holds a critical security witness.

## 2. Importance of Data Security in Financial Services

The expansion of online financial services has increased the need for robust data security, as multiple institutions share sensitive information and collaborate in complex environments. Financial computations often involve multiple parties, each with private inputs, who must securely compute functions without revealing their data, a concern critical for intrusion phishing and cheating detection [14],[15], [16]. SMPC enables such privacy-preserving computations, but challenges remain due to network asynchrony and large-scale data. High-performance SMPC protocols are therefore essential to efficiently handle massive transactions while ensuring system integrity and trustworthiness [8].

### 2.1. Challenges in Data Security

In financial services, liabilities are often backed by sensitive data such as security positions and risk exposures, requiring regular compliance reporting and audits across organizations. Collateral serves to secure these liabilities through deposits, money, securities, or investment property. The financial data security landscape is complex and costly, with billions spent on protective solutions. Implementing SMPC can enable secure handling of automated reporting and monitoring requests, allowing computations on sensitive data while preserving privacy, though system design must carefully balance functionality and protection to avoid data loss or manipulation [9].

### 2.2. Relevance of SMPC

SMPC enables multiple distrustful parties to jointly compute functions on private inputs without revealing the inputs themselves. This paradigm has generated extensive academic and industrial interest, with applications in areas such as database querying, intrusion detection, financial analysis, and data mining [8]. Despite its potential, current SMPC protocols often lack efficiency and robustness for real-world settings, particularly in financial services where data privacy is critical. Emerging practices, like cross-institutional data analysis for fraud detection [17], highlight the need to adapt existing SMPC protocols or develop tailored solutions, demonstrating the technology's transformative potential for the financial sector [9].

## 3. Fundamentals of SMPC

### 3.1. Background

SMPC also known as secure function evaluation (SFE), allows a group of parties to jointly compute a public function ($F$) with input values $x_1, x_2, ., x_n$ owned by each party ($P_1, P_2, ., P_n$), while keeping the inputs values private [8]. More formally, the privacy of means that at the end of the computation, every party ($P_i$) learns only its own input value and the computation result, but nothing else.

There are several approaches to SMPC, the most famous being Yao's Garbled Circuits and Goldwasser-Micali-Wicklin (GMW) protocol. In general, SMPC supports secrecy of gives a level of security to computations, though there is a drawback, a vast increase of the complexity of these computations in terms of memory and bandwidth usage. However, there are many applications that are simple enough and at the same time so useful that if done with some security level their impact would be substantial [18].

### 3.2. Mathematical Preliminaries

Let $G$ be the finite abelian group used as underlying group of the cryptographic scheme. In this context it is assumed $G$ is prime ordered. Let thus $p$ be the prime order of $G$. Let $g$ be a generator of $G$. Given $\alpha$ and $g$, $G^\alpha$ denotes $g^\alpha$.

In the following, for any $x \in G^\alpha$, $x_{\{i\}}, i \in \{1, \ldots, t\}$ are shares of $x$ such that $x_1 + \ldots + x_t = x$, there is the homomorphic operation $(x_1, x_2) \xrightarrow{\oplus} (y_1, y_2)$, where $y_1 = x_1 + x_2$ and $y_2 = x_2 - x_1$.

Thus, the reconstructing operation
$(y_1, y_2) \xrightarrow{Dec} x, G^\alpha$ produce $x$ if $y_1 + y_2 = x$.

On sharing generation, it is supposed that given ($x \in G^\alpha, S_1, \ldots, S_t$) are in $\mathbb{Z}_p$ and $y_i = S_i \bmod p$, where $i \in \{1, \ldots, t\}$ and $S = S_1 + S_2 \ldots + S_t$, then there will be $S \equiv 0 \ (mod\ p)$.

Given the context, it is assumed that all parties share in a trusted way ($S_{\{1,P_1\}}, S_{\{2,P_2\}}, \ldots, S_{\{t,P_n\}}$).

### 3.3. Mathematical Foundations

SMPC can guarantee the joint computation of functions over distributed data without revealing the raw data or the intermediate results to all but authorized parties [9]. In this system, a trusted party like a cryptographic module may be integrated and act as a mediator between parties. Each party inputs a secret into this mediator, which computes a value based on the aggregate of inputs while releasing only the computation result. The trusted party approach can be extended to the model without a trusted party. However, in SMPC, party corruption occurs. Before computing a binary addition, a random number is generated as a secret share. In this scenario, shares of party 1 are ⟨x⟩, ⟨y⟩, while the shares held by party 2 are ⟨x'⟩, ⟨y'⟩.

The following secret-sharing properties are required in SMPC:
(1) Randomness Parameter. The used random number is uniformly distributed in [0,1].
(2) Additive Property. For every party $P_i$, $⟨s⟩_i+⟨r⟩_i$ = r. (3) Secured property. The disclosed information reveals nothing about s to the adversary [11]. Given these properties, a binary addition can be computed. In mathematical terms, the share of the result R equals the sum of each party shares:

⟨R⟩=⟨x⟩+⟨y⟩=⟨x⟩+⟨y⟩+⟨x'⟩+⟨y'⟩=R+⟨r⟩$_1$+⟨r⟩$_2$.

### 3.4. Protocols and Algorithms

SMPC is a model in which stakeholders collaboratively compute functions on private data without revealing their inputs. Originating from Andrew Yao's 1982 concept, SMPC protocols are mainly classified into two types: evaluate-and-compile protocols, which securely evaluate functions and compile larger computations from them, and gate-by-gate protocols, which execute secure computations for each individual gate in a circuit [11].

## 4. Use Cases of SMPC in Financial Services

SMPC is a cryptographic method that allows multiple parties to jointly compute functions over private inputs without revealing them. In financial services, SMPC enables secure collaboration among institutions and regulators for tasks like fraud detection, risk assessment, and market analysis, while ensuring data privacy and regulatory compliance [19]. It supports innovation by allowing organizations to leverage shared intelligence without compromising confidentiality [20].

### 4.1. Fraud Detection

SMPC enhances fraud detection in financial services by enabling multiple parties to collaboratively analyze transactions without revealing sensitive data. This approach allows firms and telecom operators to detect fraud efficiently, controlling false alarms while maintaining high detection accuracy, all within manageable computational costs [21].

### 4.2. Risk Assessment

Risk SMPC enables collaborative risk assessment among financial institutions by allowing joint computation of measures like Value-at-Risk and Expected Shortfall without revealing individual portfolios [21]. It supports both fixed-length and variable portfolios and can incorporate additional risk factors for more comprehensive analysis, while preserving data privacy through secure computations such as matrix-matrix operations on shared secrets [11].

### 4.3. Market Analysis

SMPC allows banks and other financial institutions to collaboratively analyze data—such as transaction patterns, credit risk, and fraud detection—while maintaining the confidentiality of each party's inputs [21]. It ensures that private data remains undisclosed, outputs are shared only with entitled parties, and the protocol remains secure even against coalitions of malicious participants [11]. SMPC thus provides a practical cryptographic solution for secure, distributed computation in financial services and other domains like e-health and auctions [9].

## 5. Implementation Challenges and Solutions

The adoption of SMPC in financial services faces several practical challenges. Scalability is critical, as solutions must handle large volumes of client data and operate efficiently in real-time environments [8]. Performance overhead is another concern, since privacy-preserving computations can be resource-intensive, necessitating optimization or specialized hardware. Additionally, integrating SMPC with existing legacy systems poses difficulties, requiring architectures that can function within established financial infrastructures without complete overhauls. Addressing these issues is essential for broader implementation in the sector [22].

### 5.1. Scalability Issues

The scalability of SMPC is addressed through various approaches to handle fluctuating workloads, from hundreds to thousands of transactions per second. Strategies include horizontal scaling, network speed enhancements, parallel execution across machines, semi-honest assistant servers, and replicated secret sharing [22]. Evaluations of protocol rounds reveal challenges in maintaining high throughput while controlling latency. Proposed solutions, such as transaction batching, cascading architectures, and geographically replicated networks, aim to improve SMPC's performance for privacy-preserving financial services [9].

### 5.2. Performance Overhead

SMPC can incur high overhead in multi-party business environments due to varying infrastructures, but private or provider-managed setups make it more manageable. Performance overhead can be reduced through strategies such as establishing baseline trust and operational standards in business-to-business cooperation, as well as optimizing protocol efficiency. There is a trade-off between the number of messages exchanged and the computational operations [23], both of which impact overall execution time while maintaining security [18].

### 5.3. Integration with Legacy Systems

Large financial institutions face challenges integrating SMPC with existing infrastructures due to regulatory, operational, and cultural constraints, as well as high initial costs. Full SMPC systems, where no plaintext is exposed, may be impractical, but modular "a la carte" solutions can enhance existing single-point systems. SMPC can be implemented using various hardware or standard CPUs, coordinated via a central controller, allowing flexibility across technologies and vendors. Successful deployment requires careful management to ensure data control and computational accuracy, highlighting the need for a thorough understanding of SMPC integration [24].

## 6. Case Studies and Real-World Applications

This section highlights real-world implementations of SMPC in banking, insurance, and investment management by companies like Enveil, QEDIT, ZKProof, ZKTube, and Hive Computing. Enveil enables secure data analysis and collaboration using SMPC and homomorphic encryption without exposing raw data. QEDIT applies SMPC with zero-knowledge proofs for scalable, privacy-preserving on-chain transactions and compliance solutions. ZKProof promotes standardized adoption of ZKPs for identity, compliance, voting, and cryptocurrency. ZKTube provides practical applications and tutorials for ZKP-based products, while Hive Computing uses SMPC for private equity benchmarking and mitigating risks like front-running, ensuring transaction privacy and accurate asset pricing [25].

### 6.1. Banking Sector

SMPC is widely applied in financial services, a sector spending roughly $100 billion annually on cybersecurity, fraud prevention, and data protection. It enables banks to securely collaborate on credit ratings, risk exposure, and loan defaults without disclosing sensitive data. SMPC also allows institutions to analyze budgets and portfolios privately, supporting risk and return assessments. Early real-world implementations include fraud detection by Mitre Corporation and the FBI, and stock analysis for NASDAQ by Enron and GE [9].

### 6.2. Insurance Industry

Insurance In the insurance sector, sensitive data from customers, insurers, and healthcare providers are crucial for operations like premium assessment and fraud detection [26], yet parties are reluctant to share it. SMPC enables these stakeholders to collaboratively compute outcomes without revealing their private data. For instance, in premium assessment, SMPC allows insurers to calculate fair and accurate premiums based on customer risk without exposing underlying sensitive information, effectively addressing issues like adverse selection while maintaining privacy [8]

### 6.3. Investment Management

In investment management, strategies rely on large, sensitive datasets and complex mathematical models to guide asset allocation, risk assessment, and hedging [9]. SMPC enables multiple parties to collaboratively manage investments without revealing proprietary data or algorithmic details. While this preserves confidentiality and control over sensitive information, it carries inherent financial risks, as losses depend on the collective actions of both trustworthy and potentially dishonest participants [11].

## 7. Future Trends and Research Directions

Future research in SMPC should focus on enhancing protocol efficiency to handle large-scale, computation-intensive tasks, potentially through sub-linear complexity designs and optimized use of computing resources [9].. Additionally, exploring interoperability with other cybersecurity technologies is important, aiming to develop hybrid architectures that combine SMPC with existing measures while maintaining robust security without introducing new vulnerabilities [8]. Moreover, emerging domains such as navigation systems [27], [28] for autonomous mobile robots [29], [30], [31] and recommendation systems [32] and speech recognition platforms [33], [34], [35], [36] in financial services could leverage SMPC to securely process sensitive data across distributed nodes without compromising user privacy or operational safety.

### 7.1. Enhancements in SMPC Protocols

Future enhancements in SMPC protocols aim to improve efficiency, security, fairness, inclusion, and anonymity, with a focus on reducing communication complexity to lower execution times. Approaches include designing protocols with fewer messages, leveraging hardware acceleration, and enabling parallel execution, all of which could expand SMPC's applicability to large-scale financial and big data applications [18].

### 7.2. Interoperability with other Security Technologies

SMPC has strong potential for integration with other security technologies [37], including Federated Learning [38], [39], transformer-based models [40], Differential Privacy [8], and homomorphic encryption. Future research should explore combined approaches, such as using SMPC with FL and DP or applying SMPC on homomorphically encrypted data, to optimize performance and privacy. Additionally, extending SMPC to support Predictive Model Markup Language (PMML) could enable secure multi-party execution of predictive models across industries, while reducing reliance on fully trusted parties and broadening applicability beyond credit scoring [41].

## 8. Conclusion and Summary

The finance sector is transitioning toward greater transparency and privacy preservation, driven by innovations like smart contracts and decentralized finance (DeFi). This shift enables automated, self-enforcing transactions without human intervention, but raises challenges in protecting sensitive, non-standardized financial data.

SMPC protocols, enhanced with public encryption, are efficient and scalable, making them suitable for complex financial applications. They hold significant potential in retail finance for stock trading, real-time pricing, risk management, and creating innovative, lower-risk financial products, including solutions for crises like subprime mortgages and emerging markets such as carbon trading.